\documentclass{lmcs}
\pdfoutput=1
\usepackage[utf8]{inputenc}

\usepackage{lastpage}
\lmcsdoi{22}{2}{22}
\lmcsheading{}{\pageref{LastPage}}{}{}%
{Jul.~02,~2025}{May~25,~2026}{}


\usepackage{mystyle,mymath,latexsym,amssymb,stmaryrd}
\usepackage{amsmath,color}

\allowdisplaybreaks

\def\intmapsto{\hra}
\def\Bar{\overline}
\let\TmpA=\[
\let\TmpB=\]
\def\[{\TmpA\begin{array}{l}}
\def\]{\end{array}\TmpB}

\newcommand{\semb}[1]{[\![ #1 ]\!]}
\newcommand{\SLN}{\mathsf{SLN}}
\newcommand{\Mult}{\mathsf{Mult}}
\newcommand{\Add}{\mathsf{Add}}
\newcommand{\PA}{\mathsf{PA}}
\newcommand{\cN}{\mathcal{N}}
\newcommand{\bN}{\mathbb{N}}
\newcommand{\Var}{{\rm Vars}}

\newcommand{\Dom}{\mathsf{Dom}}
\newcommand{\hra}{\hookrightarrow}
\newcommand{\Hab}{H_\mathsf{Add1}}
\newcommand{\Hai}{H_\mathsf{Add2}}
\newcommand{\Hmb}{H_\mathsf{Mult1}}
\newcommand{\Hmi}{H_\mathsf{Mult2}}
\newcommand{\Hineqa}{H_\mathsf{Ineq1}}
\newcommand{\Hineqb}{H_\mathsf{Ineq2}}

\newcommand{\Ineq}{\mathsf{Ineq}}
\newcommand{\FV}{\mathsf{FV}}

\begin{document}
\title[Encoding Peano Arithmetic in Separation Logic]{Encoding Peano Arithmetic in a Minimal Fragment of Separation Logic}
%
%\titlerunning{Abbreviated paper title}
% If the paper title is too long for the running head, you can set
% an abbreviated paper title here
%

\author[S.~Ito]{Sohei Ito\lmcsorcid{0000-0002-7029-664X}}[a]
\author[M.~Tatsuta]{Makoto Tatsuta}[b]

\address{Nagasaki University, Japan}	%optional
% write emails for all authors having that affiliation
\email{s-ito@nagasaki-u.ac.jp}  %optional

\address{Toho University, Japan}	%optional
\email{tatsuta008@gmail.com}  %optional

\begin{abstract}
This paper investigates the expressive power of a minimal fragment of
separation logic extended with natural numbers. Specifically, it
demonstrates that the fragment consisting solely of the intuitionistic
points-to predicate, the constant 0, and the successor function is
sufficient to encode all $\Pi^0_1$ formulas of Peano Arithmetic (PA). The
authors construct a translation from PA into this fragment, showing that
a $\Pi^0_1$ formula is valid in the standard model of arithmetic if and only
if its translation is valid in the standard interpretation of the
separation logic fragment. This result implies the undecidability of
validity in the fragment, despite its syntactic simplicity. The
translation leverages a heap-based encoding of arithmetic
operations—addition, multiplication, and inequality—using structured
memory cells. The paper also explores the boundaries of this encoding,
showing that the translation does not preserve validity for $\Sigma^0_1$
formulas. Additionally, an alternative undecidability proof is presented
via a reduction from finite model theory. Finally, the paper establishes
that the validity problem for this fragment is $\Pi^0_1$-complete,
highlighting its theoretical significance in the landscape of logic and
program verification.
\end{abstract}

\maketitle

\section{Introduction}\label{sec:introduction}

Separation logic has proved to be both theoretically robust and
practically effective for verifying heap-manipulating programs
\cite{ohearn11jacm,ohearn19cacm}, owing to its concise representation of
memory states. However, when verifying software systems that involve
numerical computations, it becomes necessary to extend separation logic
with arithmetic. This extension raises fundamental questions about the
decidability of validity in such systems under standard interpretations
of numbers, as logical frameworks with decidable validity are generally
more suitable for software verification.

Presburger arithmetic, known for its decidable validity, offers a
promising candidate for integration with separation logic. One might
expect that combining a decidable fragment of separation logic with
Presburger arithmetic would yield a decidable system. Contrary to this
expectation, we show that even a minimal extension of separation logic
with arithmetic can lead to undecidability.

In this paper, we study a minimal fragment of separation logic—denoted
$\SLN$—that includes only the intuitionistic points-to predicate
($\hra$), the constant 0, and the successor function $s$. We prove that
this fragment is expressive enough to simulate all $\Pi^0_1$ formulas of
Peano Arithmetic ($\PA$). Our main result is a representation theorem
establishing a translation from $\Pi^0_1$ formulas in $\PA$ to $\SLN$
formulas that preserves both validity and non-validity under their
respective standard interpretations. As a corollary, we derive the
undecidability of validity in $\SLN$. It is surprising that such a weak
fragment of separation logic becomes so expressive merely by adding $0$
and $s$, especially considering that the fragment containing only
$\intmapsto$ is known to be decidable with respect to validity under the
standard interpretation \cite{brochenin}.

The core technique in proving our representation theorem involves
constructing an operation table for addition, multiplication, and
inequality within a heap. This allows us to eliminate expressions such
as $x + y = z$, $x \times y = z$, and $x \le y$ by referencing either
the value $z$ or the truth of the inequality $x \le y$. To encode these
operations, we use consecutive heap cells containing entries of the form
$0, x+3, y+3, x+y+3$ for addition, $1, x+3, y+3, x \times y+3$ for
multiplication, and $2, x+3, y+3$ for inequality. Here, the constant $3$
serves as an offset, and the tags $0$, $1$, and $2$ distinguish between
the three operations. To define the translation formally, we introduce
the notion of a normal form for bounded formulas in Peano Arithmetic.

While our translation can be extended to arbitrary $\PA$ formulas, the
representation theorem does not hold beyond the $\Sigma^0_1$ class. 
We provide a counterexample involving a $\Sigma^0_1$ formula to illustrate this
limitation. 

Our result shows that
discussion about properties described by $\Pi^0_1$ formulas such as
consistency of logical systems and
strong normalization properties for reduction systems
in Peano arithmetic can be simulated in
the separation logic with numbers.
The undecidability of validity in the separation logic with numbers
itself can be proved in a simpler way, by using
a similar idea to \cite{ohearn01}. 
We will also give a proof in that way.

Our representation theorem implies that the validity problem in $\SLN$
is $\Pi^0_1$-hard, establishing the lower bound of its complexity. For
the upper bound, we show that the problem belongs to $\Pi^0_1$ by
proving (1) that the model-checking problem in $\SLN$ is decidable, and
(2) that validity in $\SLN$ can be expressed as a $\Pi^0_1$ formula
using (1). Together, these results establish that the validity problem
in $\SLN$ is $\Pi^0_1$-complete.

There are several undecidability results concerning validity in
separation logic, some of which rely on translation techniques. For
instance, separation logic with the 1-field points-to predicate and the
separating implication is known to be undecidable with respect to
validity \cite{brochenin}. Similarly, separation logic with the 2-field
points-to predicate has also been shown to be undecidable
\cite{ohearn01}, with the proof relying on a translation from
first-order logic with a single binary relation into separation logic
with the 2-field points-to predicate.

On the other hand, there are some decidability results. Separation logic
with the 1-field points-to predicate, when the separating implication is
excluded, is known to be decidable \cite{brochenin}. Additionally,
the quantifier-free separation logic has been shown to be decidable
\cite{calcagno05}, with the proof involving a translation into
first-order logic with an empty signature.

To the best of our knowledge, there is no existing work that translates
any fragment of arithmetic into such a weak form of separation logic
containing only ${\intmapsto, 0, s}$.

When restricted to symbolic heaps in separation logic with arithmetic or
inductive definitions, several decidability results have been
established. These include symbolic heap entailment with Presburger
arithmetic \cite{tatsuta16}, bounded-treewidth symbolic heap entailment
\cite{iosif13}, symbolic heap entailment with cone inductive definitions
\cite{tatsuta19,nakazawa20}, symbolic heap entailment with lists
\cite{ohearn04,ohearn05,cook,antonopoulos}, and symbolic heap entailment
with Presburger arithmetic, arrays, and lists
\cite{DBLP:journals/lmcs/KimuraT21}. The satisfiability problem for
symbolic heaps with general inductive predicates is also known to be
decidable \cite{DBLP:conf/csl/BrotherstonFPG14}.

However, even within symbolic heaps, relaxing certain conditions can
lead to undecidability. For instance, symbolic heap entailment with
unrestricted inductive definitions \cite{iosif13}, and symbolic heap
entailment with bounded-treewidth inductive definitions and implicit
existentials \cite{tatsuta-kimura15}, are both known to be
undecidable. A comprehensive study on the decidability of symbolic heaps
is provided in \cite{DBLP:conf/lpar/KatelaanZ20}.

Another fragment of separation logic with arithmetic has been
shown to have a decidable satisfiability problem. Specifically, when the
fragment includes inductive predicates that capture both shape and
arithmetic properties, satisfiability remains decidable provided the
arithmetic constraints can be expressed as semilinear sets—which are
themselves decidable in Presburger arithmetic
\cite{DBLP:conf/cav/LeT0C17}.

A recent study has established the undecidability of the entailment
problem in separation logic with inductively defined spatial predicates
when certain forms of theory reasoning are permitted
\cite{DBLP:journals/ipl/EchenimP23}. This includes reasoning over
theories involving the successor function and numerical values. In
contrast, our result focuses on a significantly more restricted setting:
we allow only the 1-field intuitionistic points-to predicate as the
spatial component, which is far less expressive. Nevertheless, we prove
that even within this highly limited fragment, the logic becomes
undecidable.

This paper is organized as follows:
Section 2 defines Peano arithmetic.
Section 3 introduces the $\SLN$ fragment and its semantics.
Section 4 presents the translation from normal $\PA$ formulas to $\SLN$ and proves the preservation properties.
Section 5 defines an auxiliary translation to normal form and proves
the main theorem on the preservation of the translation from $\Pi^0_1$ formulas 
to $\SLN$.
Section 6 provides an alternative proof of undecidability of validity in $\SLN$.
Section 7 establishes $\Pi^0_1$-completeness of $\SLN$.
Section 8 concludes.

This paper extends our previous conference publication
\cite{DBLP:conf/fscd/ItoT24} by enriching the related work, clarifying
technical details, and adding the new result on $\Pi^0_1$-completeness.

\section{Peano arithmetic}
\label{sec:PA}

In this section, we define Peano arithmetic $\PA$ and its standard model.

Let $\Var = \{x,y,\ldots \}$ be the set of variables.
The \emph{terms of $\PA$} are defined by:
\[
 t ::= x ~|~ 0 ~|~ s(t) ~|~ t+t ~|~ t \times t.
\]

The \emph{formulas of $\PA$} are defined by:
\[
A ::= t = t ~|~ t \le t ~|~ \neg A ~|~ A \wedge A ~|~ A \vee A ~|~ \exists x A ~|~ \forall x A.
\]
We will write $A \to B$ for $\neg A \vee B$.

We write $s^n(t)$ for $\overbrace{s(\ldots(s}^{n}(t))\ldots)$.
We use the abbreviation $\Bar{n} = s^n(0)$.
We write $A[x:=t]$ for the formula obtained by capture-free substitution of $t$ for $x$ in $A$.

Let $\cN$ be the \emph{standard model} of $\PA$, namely, its universe $|\cN|$ is $\bN = \{0,1,2,\ldots\}$, $0^\cN$=0, $s^\cN(x) = x+1$, $+^\cN(x,y) = x + y$, 
$\times^\cN(x,y) = x \times y$, $(\le)^\cN(x,y)$ iff $x \le y$.
Let $\sigma:\Var \to \bN$ be a \emph{variable assignment}.
We extend $\sigma$ to terms in a usual way.
We write $\sigma[x:=n]$ for the variable assignment that assigns $n$ to $x$ and $\sigma(y)$ to $y$ other than $x$.

We write $\sigma \models A$ when $A$ is true in $\cN$ under the variable assignment $\sigma$.
This relation is defined in a usual way.
If $\sigma \models A$ for every variable assignment $\sigma$,
$A$ is defined to be \emph{valid}.
If $A$ does not contain free variables, $A$ is called \emph{closed}.

A formula $\forall x \le t.  A$ is an abbreviation of $\forall x(x \le t \to A)$, where $t$ does not contain $x$.
A formula $\exists x \le t. A$ is an abbreviation of $\exists x(x \le t \wedge A)$, where $t$ does not contain $x$.
We call $\forall x \le t$ and $\exists x \le t$ \emph{bounded quantifiers}.
A formula $A$ is defined to be \emph{bounded} if every quantifier in $A$ is bounded.
If $A \equiv \forall x B$ and $B$ is bounded, $A$ is called a \emph{$\Pi^0_1$ formula}.

\section{Separation logic with numbers $\SLN$}
\label{sec:SL}

In this section, we define a small fragment $\SLN$ of separation logic
with numbers.
We will also define the standard interpretation of $\SLN$.

Let $\Var = \{x,y,\ldots \}$ be the set of variables.
The \emph{terms of $\SLN$} are defined by:
\[
 t ::= x ~|~ 0 ~|~ s(t).
\]

The \emph{formulas of $\SLN$} are defined by:
\[
 A ::= t = t ~|~  t \hra t ~|~ \neg A ~|~ A \wedge A ~|~ 
A \vee A ~|~ \exists x A ~|~ \forall x A.
\]
We will write $A \to B$ for $\neg A \vee B$.

The predicate $t_1 \hra t_2$ is the intuitionistic points-to predicate and means that there is some cell of address $t_1$ which contains $t_2$ in the heap.

We use the same abbreviation $\Bar{n}$ and substitution $A[x:=t]$ as in $\PA$.
For simplicity, we write $(t \hra t_1, \ldots, t_n)$ for $t \hra t_1 \wedge \ldots \wedge s^{n-1}(t) \hra t_n$.
We sometimes write only one quantifier for consecutive quantifiers in a usual way like $\forall xy \exists zw$ for $\forall x \forall y \exists z \exists w$.

Now we define the \emph{standard interpretation} $\semb{ \cdot }$ of $\SLN$.
We use $\bN$ for both the sets of addresses and values.
Let $\semb{0}=0, \semb{s}(x) = x+1$.
Let $\sigma:\Var \to \bN$ be a \emph{variable assignment}.
The extension of $\sigma$ to terms and the variable assignment $\sigma[x:=n]$ are defined similarly to those in $\PA$.
A \emph{heap} is a finite function $h: \bN \to_\mathrm{fin} \bN$.
A heap represents a state of the memory.

For a formula $A$ of $\SLN$, we define $\sigma, h \models A$ by:
\[
\begin{array}{lcl}
\sigma, h \models t_1 = t_2 & \text{iff} & \sigma(t_1) = \sigma(t_2), \\
\sigma, h \models t_1 \hra t_2 & \text{iff} & h(\sigma(t_1)) = \sigma(t_2), \\
\sigma, h \models \neg A & \text{iff} & \sigma, h \not \models A, \\
\sigma, h \models A_1 \wedge A_2 & \text{iff} & \sigma, h \models A_1 \text{ and } \sigma, h \models A_2, \\
\sigma, h \models A_1 \vee A_2 & \text{iff} & \sigma, h \models A_1 \text{ or } \sigma, h \models A_2, \\
\sigma, h \models \exists x A & \text{iff} & \text{for some } n \in \bN, \sigma[x:=n], h \models A, \\
\sigma, h \models \forall x A & \text{iff} & \text{for all } n \in \bN, \sigma[x:=n], h \models A.
\end{array} 
\]

$\sigma, h \models A$ means that $A$ is true under the variable assignment $\sigma$ and the heap $h$.
A formula $A$ is defined to be \emph{valid} if $\sigma, h \models A$ for all $\sigma$ and $h$.
If a formula does not contain atoms $t \hra u$, the truth of the formula does not depend on heaps.

The notion of validity defined in this section refers to validity under
the standard interpretation of $\SLN$, which differs from the
conventional notion of validity in separation logic. This difference
arises because, in the conventional definition, the interpretation
depends on the set of addresses.

Our fragment $\SLN$ contains only equality, the intuitionistic points-to
predicate ($t_1 \hra t_2$), the constant $0$, successor function $s$,
Boolean connectives, and first-order quantifiers.  It omits separating
conjunction $(*)$ and magic wand $(- \! *)$.  Thus $\SLN$ is essentially
first-order logic over a finite partial function $h: \bN
\to_\mathrm{fin} \bN$ enriched with a single spatial atom.

We retain the term ``separation logic fragment'' because the semantics
of $\hra$ is the standard {\sf SL} heap-cell predicate, and our results
delineate how adding only arithmetic symbols $0,s$ to this minimal
spatial core already suffices to represent all $\Pi^0_1$ $\PA$
sentences.  We contrast our setting with classical undecidability and
decidability borders for {\sf SL} fragments in Section~\ref{sec:introduction} related work (e.g., with/without $*$, or with
multi-field points to predicate).

\section{Translation of Normal Formulas in $\PA$ into $\SLN$}
\label{sec:PA2SL}

In this section, we define the translation $( \cdot )^\circ$ of bounded formulas in $\PA$ to formulas in $\SLN$, and prove that the translation preserves the validity and the non-validity.

The key of the translation is to keep an operation table
for addition, multiplication and inequality in a heap, and
a resulting formula in $\SLN$ refers to the table instead
of using the addition, multiplication and inequality symbols.
To state that a heap keeps the operation table,
we will use a table heap condition.
For proving the preservation of the translation,
we will use a simple table heap, which is a heap that
contains all the operation entries of some size.
Since the table in a heap is finite,
to estimate the necessary size of the operation table for translating a given
formula,
we will use the upper bound of arguments in the formula.

We will first define normal form of a bounded formula in $\PA$,
which we will translate into a formula in $\SLN$.
Next we will define
a table heap condition, which guarantees that a heap has an operation table
for addition, multiplication and inequality.
Then we will define the translation of a normal formula in $\PA$
into a formula in $\SLN$.
Then we will define a simple table heap and
the upper bound of arguments in a formula.
Finally we will prove the preservation of the translation.

We write $\exists (x = t) A$ for an abbreviation of $\exists x(x = t \wedge A)$, where $t$ does not contain $x$.

Our translation is defined only for normal formulas.
This does not lose the generality since any bounded formula can
be transformed into a normal formula, as will be shown in Section 5.
In a normal formula, $+$ and $\times$ appear only in
$t$ of $\exists(x=t)$. Moreover, this $t$ is of the form $a+b$ or $a \times b$ where $a,b$ do not contain $+$ or $\times$.

\begin{defi}[Normal form]
\emph{Normal forms} of $\PA$ are given by $A$ in the following grammar:
\begin{align*}
 A ::= B ~|~ \forall x \le t . A ~|~ \exists x \le t . A ~|~ \exists(x=t) A
\end{align*}
satisfying the following conditions: (1) $B$ is a disjunctive normal form of a formula in $\PA$ without quantifiers, $+$, $\times$, and formulas of the form $\neg (t \le u)$,
(2) each $t$ in $\forall x \le t$ and $\exists x \le t$ does not contain $+, \times$,
 and (3) each $t$ in $\exists(x=t)$ is of the form $a+b$ or $a \times b$ for some terms $a$ and $b$ that do not contain $+$ or $\times$.
\end{defi}

The table heap condition is defined as the formula $H$ in the next definition.
It guarantees that a heap that satisfies $H$ contains a
correct operation table for $+$, $\times$ and $\le$.
The formulas $\Add$, $\Mult$ and $\Ineq$ in the following definition
 refer to the operation table when
a heap satisfies the table heap condition.
The normal formula enables us to represent each occurrence of $+$, $\times$ 
and $\le$ by $\Add$, $\Mult$ and $\Ineq$, respectively.
We will write $[t]$ for $s^3(t)$ for readability, since the offset is 3.

\begin{defi}[Table Heap Condition]\label{def:H}
$H$, $\Add(x,y,z)$, $\Mult(x,y,z)$ and $\Ineq(x,y)$ are the formulas defined by:
 \begin{align*}
\Hab & \equiv \forall ay ( (a \hra 0, [0], [y]) \to s^3(a) \hra [y]), \\
\Hai & \equiv \forall axy (
 (a \hra 0, [s(x)], [y]) \\
&\qquad \to \exists bz ( (b \hra 0, [x], [y], [z]) \wedge s^3(a) \hra [s(z)]) 
), \\
\Hmb& \equiv \forall ay ( (a \hra \Bar{1}, [0], [y]) \to s^3(a) \hra [0]), \\
\Hmi&\equiv \forall axy ( (a \hra \Bar{1}, [s(x)], [y]) \to \exists bz ( (b \hra \Bar{1}, [x], [y], [z]) \wedge \\
&\qquad \qquad \exists cw(
 (c \hra 0, [z], [y], [w]) \wedge s^3(a) \hra [w]
)
)
), \\
\Hineqa &\equiv \forall a x y ((a \hra \Bar{2}, [s(x)], [y])
\to \exists z b(y = s(z) \wedge (b \hra \Bar{2}, [x], [z])) ), \\
\Hineqb &\equiv \forall a x y ((a \hra \Bar{2}, [s(x)], [y] ) \to \exists b (b \hra \Bar 2, [x], [y]) ), \\
H &\equiv \Hab \wedge \Hai \wedge \Hmb \wedge \Hmi \wedge \Hineqa \wedge \Hineqb, \\
\Add(x,y,z) &\equiv \forall a ((a \hra 0, [x], [y]) \to s^3(a) \hra [z]), \\
\Mult(x,y,z) &\equiv \forall a ((a \hra \Bar{1}, [x], [y]) \to s^3(a) \hra [z]), \\
\Ineq(x,y) &\equiv \exists a (a \hra \Bar{2}, [x], [y]).
\end{align*}
\end{defi}

The formula $H$ forces a heap to have a table that contains results of addition, multiplication and inequality for some natural numbers.
Each entry for addition and multiplication consists of four cells, and each entry for inequality consists of three cells.
If the first cell contains 0, then the entry is for addition.
If the first cell contains 1, then the entry is for multiplication.
If the first cell contains 2, then the entry is for inequality.
The second and third cells of an entry represent arguments of addition, multiplication or inequality.
The entries for $+, \times$ have the forth cells, which 
contain the results of addition or multiplication.
For inequality, if there is an entry for two arguments $x$ and $y$, 
then $x \le y$ holds.
Since 0, 1, and 2 serve as reserved tags identifying the operation (addition, multiplication, or inequality), the arguments and results are encoded by adding an offset of 3.
Any fixed offset greater than $2$ would work to avoid collisions with the tags. We use 3 for concreteness.
The definition of $H$ uses the following inductive definitions of addition and multiplication: $s(x) + y = s(x+y)$ and $(x+1) \times y = x \times y + x$.
The formulas $\Hab$ and $\Hmb$ force the base cases of addition and multiplication, respectively, and $\Hai$ and $\Hmi$ force the induction steps of addition and multiplication, respectively.
$\Hineqa$ means that if there is an entry for $x+1 \le y$, then the entry for $x \le y-1$ exists in the heap.
$\Hineqb$ means that if there is an entry for $x+1 \le y$, then the entry for $x \le y$ exists in the heap.

The conjuncts of $H$ enforce coherence of existing
operation-table entries rather than totality. If a heap contains an
entry tagged as $\Add/\Mult/\Ineq$ for arguments $(x,y)$, then the
successor cells must contain the correct outputs (Lemma~\ref{lem:H}).
If no matching entry occurs, the corresponding implication in $H$ is
trivially true.  Hence $H$ admits any finite consistent partial
operation graph that is closed under the predecessor/step conditions
spelled out in $\Hab, \Hai, \Hmb, \Hmi, \Hineqa, \Hineqb$.  This
permissiveness is crucial for our $\Pi^0_1$ preservation (Lemmas~\ref{lem:PA2hn}-\ref{lem:1}). Validity is required over all heaps, so if
some heaps are ``too small'', the translated formulas become trivially
true there, while suitable ``large'' heaps (e.g., the simple tables
$h_n$ defined in Definition~\ref{def:h_n}) realize the intended arithmetic and force correctness.  By
monotonicity of arithmetic facts, truth then aligns across all
heaps.  In contrast, this permissiveness breaks preservation for
$\Sigma^0_1$.  The witness may demand the presence of a specific table
entry that small heaps can avoid (Proposition~\ref{prop:Sigma-0-1}).

We will show that the formula $H$ actually forces the heap to have a correct table for addition, multiplication and inequality (the claims (1), (2) and (3) below).
The claim (4) below says that $H$ ensures that if a heap contains an entry for $u \le u$, then it contains all the entries for $t \le u$.

\begin{lem} \label{lem:H}
Let $\sigma$ be a variable assignment and $h$ be a heap.

(1) If $\sigma, h \models H$, $h(m)=0$, $h(m+1) = n+3$ and $h(m+2) = k+3$, 
then $h(m+3) = n + k + 3$.

(2) If $\sigma, h \models H$, $h(m)=1$, $h(m+1) = n+3$ and $h(m+2) = k+3$,
then $h(m+3) = n \times k + 3$.

(3) If $\sigma, h \models H$, $h(m)=2$, $h(m+1) = n+3$, $h(m+2) = k+3$,
then $n \le k$.

(4) If $\sigma, h \models H$, $\sigma(t) \le \sigma(u)$, $\sigma, h \models \Ineq(u,u)$, then $\sigma, h \models \Ineq(t,u)$.
\end{lem}

\begin{proof}
(1) We will show the claim by induction on $n$.

\noindent \textbf{(Base case)}
Let $n = 0$.
Since
\begin{align*}
&h(m)=0, \\
&h(m+1)=3, \\
&h(m+2)=k+3, \\
&\sigma, h \models \Hab,  
\end{align*}
we have 
\[
\sigma[a:=m,x:=n,y:=k], h \models s^3(a) \hra [y].
\]
Hence, we have
\[
 h(m+3)= \sigma[a:=m,x:=n,y:=k]([y])=k+3 = n+k + 3. 
\]

\noindent \textbf{(Induction step)}
Let $n > 0$.
Then, $n-1 \ge 0$.
Let $\sigma' = \sigma[a:=m,x:=n-1,y:=k]$.
Since
\begin{align*}
&h(m)=0, \\
&h(m+1)=n+3, \\
&h(m+2)=k+3, \\
&\sigma, h \models \Hai,
\end{align*}
we have 
\[
\sigma', h \models \exists bz ((b \hra 0, [x], [y], [z]) \wedge s^3(a) \hra [s(z)]). 
\]
Thus, there exist $q$ and $\ell$ such that 
\[
 \sigma'[b:=q,z:=\ell], h \models (b \hra 0, [x], [y], [z]) \wedge s^3(a) \hra [s(z)].
\]
That is, 
\begin{align*}
&h(q)=0, \\
&h(q+1)=(n-1)+3, \\
&h(q+2)=k+3, \\
&h(q+3)=\ell+3, \\
&h(m+3)=\ell+4.  
\end{align*}
By induction hypothesis, we have $h(q+3) = (n-1) + k + 3 = n + k + 2$.
That is, $\ell = n+k - 1$.
Thus, we have
\[
 h(m+3)=\ell+4 = n + k -1 + 4 = n+k + 3.
\]

(2) We will show the claim by induction on $n$.

\noindent \textbf{(Base case)}
Let $n = 0$.
Since
\begin{align*}
&h(m)=1, \\
&h(m+1)=3, \\
&h(m+2)=k+3,\\
&\sigma, h \models \Hmb, 
\end{align*}
we have $\sigma[a:=m,x:=n,y:=k], h \models s^3(a) \hra [0]$.
Hence, we have 
\[
h(m+3)= \sigma[a:=m,x:=n,y:=k]([0])=3 = n \times k + 3.
\]

\noindent \textbf{(Induction step)}
Let $n > 0$.
Then, $n-1 \ge 0$.
Let $\sigma' = \sigma[a:=m,x:=n-1,y:=k]$.
Since
\begin{align*}
&h(m)=1, \\
&h(m+1)=n+3, \\
&h(m+2)=k+3, \\
&\sigma, h \models \Hmi, 
\end{align*}
we have
\[
\sigma', h \models \exists bz ((b \hra \Bar{1}, [x], [y], [z]) \wedge 
\exists cw(
(c \hra 0, [z], [y], [w]) \wedge s^3(a) \hra [w]
)
).
\]
Thus, there exist $q$ and $\ell$ such that
\[
\sigma'[b:=q,z:=\ell], h \models (b \hra \Bar{1}, [x], [y], [z]) \wedge 
\exists cw(
(c \hra 0, [z], [y], [w]) \wedge s^3(a) \hra [w]
).
\]
That is,
\begin{align*}
&h(q)=1, \\
&h(q+1)=(n-1)+3, \\
&h(q+2)=k+3,\\
&h(q+3)=\ell+3. 
\end{align*} 
So by induction hypothesis, $h(q+3) = (n-1) \times k + 3$.
Thus $\ell = (n-1) \times k$.
Furthermore, since 
\[
\sigma'[b:=q,z:=\ell], h \models \exists cw(
(c \hra 0, [z], [y], [w]) \wedge s^3(a) \hra [w]), 
\]
we have
\[
\sigma'[b:=q,z:=\ell,c:=r,w:=p], h \models 
(c \hra 0, [z], [y], [w]) \wedge s^3(a) \hra [w]
\]
for some $r$ and $p$.
That is, 
\begin{align*}
&h(r)=0, \\
&h(r+1)=\ell+3, \\
&h(r+2)=k+3, \\
&h(r+3)=p+3, \\
&h(m+3)=p+3. 
\end{align*}
By (1) of this Lemma, we have $p = \ell+k$.
With this and $\ell = (n-1) \times k$, we have $p = (n-1) \times k + k = n \times k$.
Hence, $h(m+3) = p+3 = n \times k + 3$.

(3) We will show the claim by induction on $n$.

\noindent \textbf{(Base case)}
Let $n = 0$.
We immediately have $n \le k$.

\noindent \textbf{(Induction step)}
Let $n > 0$.
Then, $n-1 \ge 0$.
Let $\sigma' = \sigma[a:=m, x:=n-1, y:=k]$.
Since
\begin{align*}
&h(m)=2, \\
&h(m+1)=n+3, \\
&h(m+2)=k+3,\\
&\sigma, h \models \Hineqa, 
\end{align*}
we have 
\[
\sigma', h \models \exists z b (y = s(z) \wedge (b \hra \Bar 2, [x], [z])).
\]
Thus, there exist $\ell$ and $q$ such that
\[
\sigma'[z:=\ell, b:=p], h \models y = s(z) \wedge (b \hra \Bar 2, [x], [z]),
\]
that is,
\begin{align*}
&h(p)=2,\\
&h(p+1)=(n-1)+3,\\
&h(p+2)=\ell + 3 = (k-1)+3. 
\end{align*}
By induction hypothesis, $n-1 \le k-1$, that is, $n \le k$.

(4) We will show the claim by induction on $\sigma(u) - \sigma(t)$.

\noindent \textbf{(Base case)}
Let $\sigma(u) - \sigma(t) = 0$, i.e. $\sigma(t)=\sigma(u)$.
By assumption, we have $\sigma, h \models \Ineq(u,u)$.
Since $\sigma(t)=\sigma(u)$, we have the claim.

\noindent \textbf{(Induction step)}
Let $\sigma(u) - \sigma(t) > 0$, i.e. $\sigma(t) < \sigma(u)$.
Since
\[
 \sigma(u) - (\sigma(t)+1) < \sigma(u) - \sigma(t),
\]
we have
\[
\sigma, h \models \Ineq (s(t), u)
\]
by induction hypothesis.
That is, $\sigma, h \models \exists a (a \hra \Bar 2, [s(t)], [u])$.
Since $\sigma, h \models \Hineqb$,
we have
\[
\sigma, h \models \exists b (b \hra \Bar 2, [t], [u]),
\]
that is, $\sigma, h \models \Ineq(t,u)$.
\end{proof}

Now we define the translation of normal formulas in $\PA$ into formulas in $\SLN$.
In the translation, $+$, $\times$ and $\le$ are replaced by $\Add$, $\Mult$ and $\Ineq$ with the table heap condition.

\begin{defi}[Translation $( \cdot )^\circ$]
Let $A$ be a normal formula in $\PA$.
We define $\SLN$ formula $(\forall x A)^\circ$ as:
\begin{align*}
 B^\circ &\equiv B^\le \text{ if $B$ is quantifier-free}, \\
 (\exists x \le t. B)^\circ &\equiv H \to \neg \Ineq(t,t) \vee \exists x (\Ineq(x,t) \wedge B^\circ), \\
 (\forall x \le t. B)^\circ &\equiv H \to \forall x (\neg \Ineq(x,t) \vee B^\circ), \\
 (\exists(x=t+u)B)^\circ &\equiv H \to \exists x (\Add(t,u,x) \wedge B^\circ), \\
 (\exists(x=t \times u)B)^\circ &\equiv H \to \exists x (\Mult(t,u,x) \wedge B^\circ), \\
 (\forall x A)^\circ &\equiv \forall x A^\circ,
\end{align*}
where $B^\le$ is obtained from $B$ by replacing each positive occurrence of $t \le u$ by $H \to \neg \Ineq(u,t) \vee t = u$.
\end{defi}

For a normal formula,
the translation computes $t+u $ by
referring to
the operation table in the current heap.
$H$ guarantees that the operation table in the heap is correct.
However, the operation table may not be sufficiently large for
computing $t+u$. 
When the operation table is not sufficiently large for
computing $t+u$,
then $\Add(t,u,x)$ returns true.
The use of $\Ineq$ and $\Mult$ is similar to $\Add$.
Note that $\Ineq(t,t)$ means that $t$ is in the operation table.
Hence $\neg \Ineq(t,t)$ means that $t$ is not in the operation table
and we define $(\exists x \le t.B)^\circ$ as true in this case.
For a non-normal formula, we just keep $\forall$ in the translation.

\begin{exa}
For a normal formula
\[
A \equiv \exists(x_1=x+s(x))\exists(x_2=x+x_1)\forall y \le x_2.
\\ \qquad
\exists(x_3=x+y)\exists(x_4=y\times x_3)\exists(x_5=x+x_4)(0 \le x_5),
\]
its translation $A^\circ$ is
\[
A^\circ \equiv H \to \exists x_1(\Add(x,s(x),x_1) \wedge 
(H \to \exists x_2 (\Add(x,x_1,x_2) \wedge
\\ \qquad
(H \to \forall y (\neg \Ineq(y, x_2) \vee 
\\ \qquad
(H \to \exists x_3(\Add(x,y,x_3) \wedge 
(H \to \exists x_4(\Mult(y,x_3,x_4) \wedge 
\\ \qquad
(H \to \exists x_5(\Add(x,x_4,x_5) \wedge 
(H \to \neg \Ineq(x_5,0) \vee 0 = x_5)
))))))))))).
\]
\end{exa}

\begin{exa}
For a normal formula
\[
B \equiv \exists(x_1=x+s(x))\exists y \le x_1.(0 \le y),
\]
its translation $B^\circ$ is
\[
B^\circ \equiv H \to \exists x_1 
(\Add(x,s(x),x_1)\wedge
 (H \to \neg \Ineq(x_1,x_1) \vee 
\\ \qquad
 \exists y 
  (\Ineq(y,x_1)\wedge
   (H \to \neg \Ineq(y,0) \vee y=0
   )
  )
 )
)
\]
\end{exa}

Our goal is to show that for any $\Pi^0_1$ formula $A$ of $\PA$, $A$ is
valid in $\PA$ if and only if $A^\circ$ is valid in $\SLN$.  Therefore,
$A^\circ$ should hold for \emph{every} heap $h$.  By the definition of
$( \cdot )^\circ$, $x=t+u$ and $x = t \times u$ are translated into $H
\to \Add(t,u,x)$ and $H \to \Mult(t,u,x)$, respectively. 
Furthermore, the formulas $\Add(t,u,x)$ and $\Mult(t,u,x)$ state that
for any address $a$, if the cells at $a+1$ and $a+2$ contain the
operands $t$ and $u$, respectively, then the cell at $a+3$ contains the
result $x$. Consequently, if a heap $h$ does not include a sufficiently
large table to store the operands for $x = t + u$ or $x = t \times u$,
the translated formulas are trivially true.
Since we demand that $A^\circ$ hold for all heaps, there is $h$ that contains a sufficiently large table.
Furthermore, if the addition and multiplication in the formula are correct in such a sufficiently large heap, they must be correct in every heap, because addition and multiplication are numeric properties and do not depend on heaps.
The same is true for inequality.
This is the key idea to prove our goal.
That is, $\sigma \models A$ if and only if $\sigma, h \models A^\circ$ for \emph{sufficiently large} $h$ if and only if $\sigma, h \models A^\circ$ for \emph{all} $h$.
We will prove them in Lemmas~\ref{lem:PA2hn} and \ref{lem:hn2forallh} later.

Since we demand that $A^\circ$ hold for all heaps, we define the translation of $t \le u$ to be $H \to \neg \Ineq(u,t) \vee t=u$ and we do not straightforwardly define it to be $H \to \Ineq(t,u)$, because $\Ineq(t,u)$ demands the heap to contain the entry for $t \le u$, which is not possible if the heap is not sufficiently large.
Furthermore, the translation of $\exists x \le t. B$ is not simply $H \to \exists x(\Ineq(x,t) \wedge B)$ but rather seemingly tricky $H \to \neg \Ineq(t,t) \vee \exists x (\Ineq(x,t) \wedge B^\circ)$.
If we adopt the simple translation, we may not be able to find $x$ such that the entry for $x \le t$ is in the heap when it is not sufficiently large.
Our idea is to let such a case be true.
Therefore, we allow the case $\neg \Ineq(t,t)$, which is true if the heap may not contain some entries for $\cdot \le t$.

First, we estimate the necessary size of the operation table for 
a given formula and a given variable assignment.
This size is defined in the next definition.
\begin{defi}
 Let $A$ be prenex and disjunctive normal form of a bounded formula in $\PA$ and $\sigma$ be a variable assignment.
We define the number $\max(\sigma, A)$ by:
\begin{align*}
& \max(\sigma, t \le u) = \max \{ \sigma(t), \sigma(u) \}, \\
& \max(\sigma, t = u) = 0, \\
& \max(\sigma, \neg B) = \max(\sigma, B) \\
& \max(\sigma, B \wedge C) = \max\{\max(\sigma, B), \max(\sigma, C) \} \\
& \max(\sigma, B \vee C) = \max\{\max(\sigma, B), \max(\sigma, C) \} \\
& \max(\sigma, \forall x \le t. B) = \max \{\sigma(t), \max(\sigma, B[x:=t]) \}, \\
& \max(\sigma, \exists x \le t. B) = \max \{\sigma(t), \max(\sigma, B[x:=t]) \}, \\
& \max(\sigma, \exists (x = t) B) = \max\{ \sigma(t), \max(\sigma, B[x:=t])
 \}.
\end{align*}
\end{defi}

$\max(\sigma,A)$ computes
the maximum in values of expressions by $\sigma$ in $A'$
where $A'$ is obtained by expanding bounded quantifiers of $A$.

\begin{exa}
By using the above definition one by one, we have
\begin{align*}
&\max(\sigma,
\exists(x_1=x+s(x))\exists(x_2=x+x_1)\forall y \le x_2.
\\ &\qquad\exists(x_3=x+y)\exists(x_4=y\times x_3)\exists(x_5=x+x_4)(0 \le x_5))
\\
&= \max\{\sigma(x+s(x)),
\max(\sigma,
\exists(x_2=x+(x+s(x)))\forall y \le x_2.
  \\ &\qquad\exists(x_3=x+y)\exists(x_4=y\times x_3)\exists(x_5=x+x_4)(0 \le x_5))\}
\\
&= \max\{\sigma(x+s(x)),
\max\{\sigma(x+(x+s(x))),
\max(\sigma,
\forall y \le x + (x+s(x)).
  \\ &\qquad\exists(x_3=x+y)\exists(x_4=y\times x_3)\exists(x_5=x+x_4)(0 \le x_5)
)\}\}
\\
&= \max\{\sigma(x+s(x)),
\max\{\sigma(x+(x+s(x))),
\max\{\sigma(x+(x+s(x))),
\\ &\qquad
\max(\sigma,
  \exists(x_3=x+(x+(x+s(x))))\exists(x_4=(x+(x+s(x)))\times x_3)
\\ &\qquad \exists(x_5=x+x_4)(0 \le x_5)
)\}\}\}
\\
&= \max\{\sigma(x+s(x)),
\max\{\sigma(x+(x+s(x))),
\max\{\sigma(x+(x+s(x))),
\\ &\qquad
\max\{\sigma(x+(x+(x+s(x)))),
\\ &\qquad
\max(\sigma,
  \exists(x_4=(x+(x+s(x)))\times (x+(x+(x+s(x)))))\exists(x_5=x+x_4)(0 \le x_5)
)\}\}\}\}
\\
&= \max\{\sigma(x+s(x)),
\max\{\sigma(x+(x+s(x))),
\max\{\sigma(x+(x+s(x))),
\\ &\qquad
\max\{\sigma(x+(x+(x+s(x)))),
\\ &\qquad
\max\{\sigma((x+(x+s(x)))\times (x+(x+(x+s(x))))),
\\ &\qquad
\max(\sigma,
  \exists(x_5=x+(x+(x+s(x)))\times (x+(x+(x+s(x)))))(0 \le x_5)
)\}\}\}\}\}
    \\ 
&= \max\{\sigma(x+s(x)),
\max\{\sigma(x+(x+s(x))),
\max\{\sigma(x+(x+s(x))),
\\ &\qquad
\max\{\sigma(x+(x+(x+s(x)))),
\\ &\qquad
\max\{\sigma((x+(x+s(x))\times (x+(x+(x+s(x))))),
\\ &\qquad
\max\{\sigma(x+(x+(x+s(x)))\times (x+(x+(x+s(x))))),
\\ &\qquad
\max(\sigma,
  0 \le x+(x+(x+s(x)))\times (x+(x+(x+s(x))))
)\}\}\}\}\}\} \displaybreak
\\
&= \max\{\sigma(x+s(x)),
\max\{\sigma(x+(x+s(x))),
\max\{\sigma(x+(x+s(x))),
\\ &\qquad
\max\{\sigma(x+(x+(x+s(x)))),
\\ &\qquad
\max\{\sigma((x+s(x))\times (x+(x+(x+s(x))))),
\\ &\qquad
\max\{\sigma(x+(x+s(x))\times (x+(x+(x+s(x))))),
\\ &\qquad
\max\{0,\sigma(x+(x+(x+s(x)))\times (x+(x+(x+s(x)))))\}\}\}\}\}\}\}
\\
&= \sigma(x)+ (3\sigma(x)+1)(4\sigma(x)+1).
\end{align*}
\end{exa}

Next, for a given size $n$, we define a heap that supports addition for
arguments up to $n^2$, and multiplication and inequality for arguments
up to $n$. We refer to this as a {\em simple table heap}.

\begin{defi}\label{def:h_n}
For a number $n$, we define a heap
$h_n$ as the heap defined by:
\begin{align*}
&h_n (x) =
\begin{cases}
	     0 & (x = 4i, i < (n^2+1)^2) \\
	     i \bmod (n^2 + 1) + 3 & (x = 4i+1, i < (n^2+1)^2) \\
	     \lfloor i / (n^2 + 1) \rfloor + 3 & (x = 4i+2, i < (n^2+1)^2) \\
	     h_n(x-2) + h_n(x-1) - 3 & (x = 4i+3, i < (n^2+1)^2) \\
	     1 & (x = c_1 +4i, i < (n+1)^2) \\
	     i \bmod (n + 1) + 3 & (x = c_1 + 4i+1, i < (n+1)^2) \\
	     \lfloor i / (n + 1) \rfloor + 3 & (x = c_1 +4i+2, i < (n+1)^2) \\
	     (h_n(x-2)-3) \times (h_n(x-1)-3) + 3 & (x = c_1 + 4i+3, i < (n+1)^2) \\
             2 & (x = c_2 + 3i, i < (n+1)^2) \\
             i \bmod (n+1) + 3 & (x = c_2 + 3i+1, i < (n+1)^2) \\
             n + 3 & (x = c_2 + 3i+2, i < (n+1)^2, \\
                   & \quad \lfloor i / (n+1) \rfloor < i \bmod (n+1)) \\
             \lfloor i / (n+1) \rfloor + 3 & (x = c_2 + 3i+2, i < (n+1)^2, \\
                   & \quad \lfloor i / (n+1) \rfloor \ge i \bmod (n+1)) \\
	     \text{undefined} & \text{otherwise}
	    \end{cases}
\end{align*}
where $c_1 = 4(n^2+1)^2$ and $c_2 = c_1 + 4(n+1)^2$.
\end{defi}

The heap $h_n$ has the operation table that has entries of $+$ for
arguments up to $n^2$ and the entries of $\times$ and $\le$ for arguments
up to $n$.  The $i$-th entry for $+$ contains the result of addition of $x
= i \bmod (n^2+1)$ and $y = \lfloor i / (n^2+1) \rfloor$, that is,
$h(4i)=0$, $h(4i+1)=x+3$, $h(4i+2)=y+3$ and $h(4i+3)=x+y+3$.  The $i$-th
entry for $\times$ contains the result of multiplication of $x = i \bmod
(n+1)$ and $y = \lfloor i / (n+1) \rfloor$, that is, $h(c_1+ 4i)=1$,
$h(c_1+ 4i+1)=x+3$, $h(c_1+ 4i+2)=y+3$ and $h(c_1+4i+3)=x \times y+3$.
The $i$-th entry for $\le$ signifies inequality of $x = i \bmod (n+1)$
and $y = \lfloor i / (n+1) \rfloor$ or $n$, where $h(c_2+ 4i)=2$,
$h(c_2+ 4i+1)=x+3$, and $h(c_2+ 4i+2)=y+3$ if $x \le y$ and $h(c_2+
4i+2)=n+3$ if $x > y$.

The next lemma shows that the simple table heap $h_n$ satisfies the table heap condition $H$.

\begin{lem} \label{lem:hnH}
For a variable assignment $\sigma$, we have $\sigma, h_n \models H$.
\end{lem}

\begin{proof}
We check each conjunct of $H$.

$\Hab$: For any $a, y$, if $h_n(a)=0, h_n(a+1)=3$, and $h_n(a+2)=y+3$, then
by Definition~\ref{def:h_n} the block starting at $a$ is the $(0,0,y)$-entry
and $h_n(a+3)=y+3$.

$\Hai$: Suppose $h_n(a)=0, h_n(a+1)=x+4, h_n(a+2)=y+3$. 
By Definition~\ref{def:h_n}, $h_n(a+3)=(x+1)+y+3$.
Since $h_n$ has a block for $(0,x,y)$ at some $b$ with $h_n(b)=0,
h_n(b+1)=x+3, h_n(b+2)=y+3, h_n(b+3)=x+y+3$, the claim holds.

$\Hmb$, $\Hmi$: Analogous, using the multiplication lines in Definition~\ref{def:h_n} and the identity $(x+1)\times y = x \times y + y$.

$\Hineqa, \Hineqb$: For $h_n(a)=2$ and arguments $(x+1,y)$, Definition
\ref{def:h_n} places entries for the predecessor pair $(x,y-1)$ (sine $y>0$ by
$x+1 \le y$) and ensures the monotone closure $(x,y)$ as required.
\end{proof}

The next lemma shows that
the truth of $t+u=v$, $t\times u=v$ and $t \le u$ in $\PA$ for the standard model is 
equivalent to the truth of
their translations in $\SLN$ for the standard interpretation
for the simple table heap.

For $t+u$,
since $n$ is greater than or equal to the arguments $t,u$,
the heap $h_n$ is sufficiently large to compute
the addition $t+u$ by referring to the operation table in $h_n$.
Hence $\Add(t,u,v)$ exactly computes $t+u=v$.
The ideas are similar for $t \times u$ and $t \le u$.

\begin{lem} \label{lem:PA2hnBase}
For $n \ge \max \{\sigma(t), \sigma(u)\}$, the following hold.

(1) $\sigma \models t+u = v$ if and only if $\sigma, h_n \models \Add(t,u,v)$.

(2) $\sigma \models t \times u = v$ if and only if $\sigma, h_n \models \Mult(t,u,v)$.

(3) $\sigma \models t \le u$ if and only if $\sigma, h_n \models \Ineq(t,u)$.
\end{lem}

\begin{proof}
(1) Only-if-direction: Since $n \ge \max
\{\sigma(t), \sigma(u)\}$, by the definition of $h_n$, there exists $p$
such that 
\[
\sigma[a:=p],h_n \models (a \hra 0, [t], [u]).
\]
That is,
\begin{align*}
&h_n(p)=0, \\
&h_n(p+1)=\sigma(t)+3, \\
&h_n(p+2)=\sigma(u)+3. 
\end{align*}
By the
definition of $h_n$, we have $h_n(p+3)=\sigma(t)+\sigma(u)+3$.  By
assumption, $\sigma(t)+\sigma(u)=\sigma(v)$.  Therefore, $h_n(p+3)=
\sigma(v)+3$.  Thus, 
\[
\sigma[a:=p],h_n \models s^3(a) \hra [v].
\]
Hence,
$\sigma, h_n \models \Add(t,u,v)$.

If-direction:
Since $n \ge \max \{\sigma(t), \sigma(u)\}$, by the definition of $h_n$, there exists $p$ such that 
\[
\sigma[a:=p], h_n\models (a \hra 0, [t], [u], [v]).
\]
Thus, 
\begin{align*}
&h_n(p)=0, \\
&h_n(p+1) = \sigma(t)+3, \\
&h_n(p+2)=\sigma(u)+3, \\
&h_n(p+3) = \sigma(v)+3.
\end{align*}
By the definition of $h_n$, we have 
\[
h_n(p+3)= (h_n(p+1)-3) + (h_n(p+2)-3) + 3.
\]
Since $h_n(p+1) = \sigma(t)+3$ and $h_n(p+2) = \sigma(u)+3$, we have 
\[
h_n(p+3)= \sigma(t) + \sigma(u) + 3. 
\]
Thus, we have $\sigma(t)+\sigma(u)=\sigma(v)$.
Hence, $\sigma \models t+u = v$.

(2) The claim can be shown similarly to (1).

(3) Only-if-direction: 
Suppose $\sigma(t) \le \sigma(u)$.
Let $i = \sigma(t) \cdot (n+1) + \sigma(u)$.
Since $n \ge \max \{\sigma(t), \sigma(u)\}$, we have $i < (n+1)^2$.
Furthermore,
\begin{align*}
 \sigma(t) &= \lfloor i / (n+1) \rfloor,\\
 \sigma(u) &= i \bmod (n+1). 
\end{align*}
For $p = 4(n^2+1)^2+4(n+1)^2+3i$, we have $h_n(p)=2$, 
$h_n(p+1)=\sigma(t)+3$ by the definition of $h_n$.
Since $\sigma(t) \le \sigma(u)$, 
we have
\[
 h_n(p+2) = \sigma(u)+3
\] 
 by the definition of $h_n$.
From this, we have $\sigma, h_n \models \exists a (a \hra \Bar 2, [t], [u])$, that is, $\sigma, h_n \models \Ineq(t,u)$.

If-direction:
Suppose $\sigma, h_n \models \Ineq(t,u)$.
Since $n \ge \max \{\sigma(t), \sigma(u)\}$, there exists $p$ such that 
\begin{align*}
&h_n(p)=2, \\
&h_n(p+1)=\sigma(t)+3, \\
&h_n(p+2)=\sigma(u)+3.
\end{align*}
By Lemma~\ref{lem:H} (3), we have $h_n(p+1) \le h_n(p+2)$, that is, $\sigma(t)\le \sigma(u)$.
\end{proof}

The next lemma shows that if $\Add$, $\Mult$ and $\neg \Ineq$ are true for a sufficiently large simple table heap, they are also true for all heaps.

In (1),
since $n$ is greater than or equal to the arguments $t,u$,
the heap $h_n$ is sufficiently large to compute
the addition $t+u$ by referring to the operation table in $h_n$.
Hence $\Add(t,u,v)$ in the left-hand side exactly computes $t+u=v$.
The right-hand side takes all heaps $h$.
When the heap $h$ is not sufficiently large, the right-hand side becomes true.
When the heap $h$ is sufficiently large, the right-hand side exactly
computes $t+u=v$.
For this reason the equivalence holds.
The ideas are similar in (2) and (3).

\begin{lem} \label{lem:hn2forallhBase}
For $n \ge \max \{\sigma(t), \sigma(u)\}$, the following hold.

(1) $\sigma, h_n \models \Add(t,u,v)$ if and only if $\sigma, h \models H \to \Add(t,u,v)$ for all $h$.

(2) $\sigma, h_n \models \Mult(t,u,v)$ if and only if $\sigma, h \models H \to \Mult(t,u,v)$ for all $h$.

(3) $\sigma, h_n \models \neg \Ineq(t,u)$ if and only if $\sigma, h \models H \to \neg \Ineq(t,u)$ for all $h$.
\end{lem}

\begin{proof}
The if-direction is obvious. We will show the only-if-direction.

(1)
Since $\sigma, h_n \models \Add(t,u,v)$ by assumption, we have $\sigma(t)+\sigma(u)=\sigma(v)$ by Lemma~\ref{lem:PA2hnBase} (1).
We fix $h$ in order to show $\sigma, h \models H \to \Add(t,u,v)$.

Case 1. If $\sigma, h \not \models H$, the claim follows trivially.

Case 2. Assume $\sigma, h \models H$.

Case 2.1 If $\sigma, h \models \forall a \neg(a \hra 0, [t], [u])$, the claim follows trivially, because $\sigma, h \models \forall a((a \hra 0, [t], [u]) \to s^3(a)\hra [u])$.

Case 2.2
Assume $\sigma, h \models \exists a (a \hra 0, [t], [u])$.
We assume 
\begin{align*}
&h(p)=0,\\
&h(p+1)=\sigma(t)+3,\\
&h(p+2)=\sigma(u)+3
\end{align*}
 for arbitrary $p$.
Since $\sigma, h \models H$, we have 
\[
h(p+3)=(h(p+1)-3)+(h(p+2)-3)+3
\]
by Lemma~\ref{lem:H} (1).
Therefore, $h(p+3)= \sigma(t) + \sigma(u) + 3$.
That is, $h(p+3)= \sigma(v) + 3$.
Thus $\sigma, h \models s^3(a) \hra [v]$.
Then, we have 
\[
\sigma[a:=p], h \models (a \hra 0, [t], [u]) \to s^3(a) \hra [v] \text{\quad for all $p$.}
\]

Hence in both cases $\sigma, h \models H \to \Add(t,u,v)$.

(2)
The claim can be shown similarly to (1) (except it uses Lemma~\ref{lem:H} (2)).

(3) 
By Lemma~\ref{lem:PA2hnBase} (3), we have $\sigma \models \neg (t \le u)$.
We fix $h$ in order to show $\sigma, h \models H \to \neg \Ineq(t,u)$.

Case 1. If $\sigma, h \not \models H$, the claim follows trivially.

Case 2. Assume $\sigma, h \models H$.
Assume $\sigma, h \models \Ineq(t,u)$ for contradiction.
Then, there is $q$ such that
\begin{align*}
&h(q)=2, \\
&h(q+1) = \sigma(t)+3, \\
&h(q+2)=\sigma(u)+3.
\end{align*}
By Lemma~\ref{lem:H} (3), we have $\sigma(t) \le \sigma(u)$, a contradiction.
\end{proof}

The next lemma says that
the truth in $\PA$ is equivalent to the truth of the translation in $\SLN$ for a large simple table heap.

Lemma~\ref{lem:PA2hnBase} already proved this statement to
atomic formulas.
The next lemma is proved by
extending it to a normal formula.

\begin{lem} \label{lem:PA2hn}
 For a normal formula $A$ in $\PA$ and $n \ge \max(\sigma, A)$, $\sigma \models A$ if and only if $\sigma,h_n \models A^\circ$.
\end{lem}

\begin{proof}
We will show the claim by induction on $A$.

Case 1. $A$ is quantifier-free. 
We will only show the cases for $A \equiv (t \le u)$
since
the cases $t=u$ and $t \neq u$ are obvious and
the cases $A \wedge B$ and $A \vee B$ follow from the induction hypothesis.
$\sigma \models t \le u$ is equivalent to 
\[
\sigma \models \neg (u \le t) \vee t = u.
\]
Since $n \ge \max \{\sigma(t), \sigma(u)\}$, by Lemma~\ref{lem:PA2hnBase} (3), $\sigma \models \neg (u \le t)$ is equivalent to 
\[
\sigma, h_n \models \neg \Ineq(u,t).
\]
Hence, $\sigma \models t \le u$ is equivalent to 
\[
\sigma, h_n \models \neg \Ineq(u,t) \vee t=u.
\]
Since $\sigma, h_n \models H$ by Lemma~\ref{lem:hnH}, $\sigma, h_n \models \neg \Ineq(u,t) \vee t=u$ is equivalent to 
\[
\sigma, h_n \models H \to \neg \Ineq(u,t) \vee t=u.
\]

Case 2. $A \equiv \exists x \le t. B$.

Only-if-direction:
By assumption, there is $k$ such that 
\[
\sigma[x:=k] \models x \le t \wedge B.
\]
That is,
\[
\sigma[x:=k] \models x \le t \text{ and } \sigma[x:=k] \models B.
\]
Thus, we have $k \le \sigma(t)$.
Since $n \ge \max(\sigma[x:=k], B)$, by induction hypothesis, 
we have
\[
 \sigma[x:=k], h_n \models B^\circ.
\]
Furthermore, by Lemma~\ref{lem:PA2hnBase} (3),
\[
\sigma[x:=k], h_n \models \Ineq(x,t).
\]
Thus, we have 
\[
\sigma[x:=k], h_n \models \Ineq(x,t) \wedge B^\circ.
\]
Hence, 
\[
\sigma[x:=k], h_n \models \neg \Ineq(t,t) \vee (\Ineq(x,t) \wedge B^\circ).
\]
Thus, we have
\[
\sigma, h_n \models \neg \Ineq(t,t) \vee \exists x(\Ineq(x,t) \wedge B^\circ).
\]

If-direction:
Suppose 
\[
\sigma[x:=k], h_n \models H \to \neg \Ineq(t,t) \vee (\Ineq(x,t) \wedge B^\circ)
\]
 for some $k$.
Since $\sigma[x:=k], h_n \models H$, we have 
\[
 \sigma[x:=k], h_n \models \neg \Ineq(t,t) \vee (\Ineq(x,t) \wedge B^\circ).
\]
Since $n \ge \max(\sigma, A) \ge \sigma(t)$, by the definition of $h_n$, 
\[
 \sigma[x:=k],h_n \models \Ineq(t,t).
\]
Thus, we have 
\[
 \sigma[x:=k], h_n \models \Ineq(x,t) \wedge B^\circ.
\]
Since $\sigma[x:=k], h_n \models \Ineq(x,t)$, by Lemma~\ref{lem:PA2hnBase} (3), 
we have $k \le \sigma(t)$.
Then, since $k \le \sigma(t) \le n$, by the induction hypothesis for $B$, 
\[
\sigma[x:=k] \models x \le t \wedge B.
\]
That is,
\[
\sigma \models \exists x \le t. B.
\]

Case 3. $A \equiv \forall x \le t. B$.
We will show the claim: For all $k$, 
\[
\sigma[x:=k] \models \neg (x \le t) \vee B \text{ if and only if } \sigma[x:=k], h_n \models \neg \Ineq(x,t) \vee B^\circ.
\]
If $k \le n$, then by Lemma~\ref{lem:PA2hnBase} (3) and the 
induction hypothesis for $B$, the claim holds.
If $k > n$, then since $k > n \ge \sigma(t)$, we have 
\[
 \sigma[x:=k] \models \neg (x \le t).
\]
On the other hand, by the definition of $h_n$, we have 
\[
\sigma[x:=k], h_n \models \neg \Ineq(x,t).
\]
Thus, the claim holds, and the original statement follows directly from it.

Case 4. $A \equiv \exists(x=t+u)B$.
$\sigma \models \exists(x=t+u)B$ is equivalent to 
\[
\sigma[x:=k] \models x=t+u$ and $\sigma[x:=k] \models B 
\]
 for some $k$.
Since $n \ge \max \{\sigma(t), \sigma(u)\}$,
by Lemma~\ref{lem:PA2hnBase} (1), $\sigma[x:=k] \models x=t+u$ is equivalent to
\[
\sigma[x:=k], h_n \models \Add(t,u,x). 
\]
Furthermore, since 
\begin{align*}
n &\ge \max(\sigma,\exists(x=t+u)B)=\max\{\sigma(t+u), \max(\sigma,B[x:=t+u])\} \\
 &\ge \max(\sigma,B[x:=t+u]) = \max(\sigma,B[x:=\Bar k]) = \max(\sigma[x:=k],B), 
\end{align*}
by induction hypothesis for $B$, $\sigma[x:=k] \models B$ is equivalent to
\[
\sigma[x:=k],h_n \models B^\circ.
\]
Therefore, $\sigma \models A$ is equivalent to
\[
\sigma[x:=k], h_n \models \Add(t,u,x) \wedge B^\circ
\]
for some $k$, which is equivalent to 
\[
\sigma, h_n \models \exists x(\Add(t,u,x) \wedge B^\circ).
\]

Case 5. $A \equiv \exists(x=y\times z) B$. This case can be shown similarly to Case 4 (except it uses Lemma~\ref{lem:PA2hnBase} (2)).
\end{proof}

The next lemma says that
for the translation of a normal formula in $\PA$,
the truth for a large simple table heap is the same as
the truth for all heaps
in the standard interpretation of $\SLN$.

\begin{lem} \label{lem:hn2forallh}
 Let $A$ be a normal formula in $\PA$ and $n \ge \max(\sigma, A)$.
Then, $\sigma, h_n \models A^\circ$ if and only if
$\sigma, h \models A^\circ$ for all $h$.
\end{lem}

\begin{proof}
The if-direction is trivial.
We will show the only-if-direction by induction on $A$.

Case 1. $A$ is quantifier-free.
We will only show the case for $A \equiv (t \le u)$
since
the cases $t=u$ and $t \neq u$ are obvious and
the cases $A \wedge B$ and $A \vee B$ follow from the induction hypothesis.
Since $\sigma, h_n \models H$ by Lemma~\ref{lem:hnH},
\[
\sigma, h_n \models H \to \neg \Ineq(u,t) \vee t=u  
\]
is equivalent to
\[
\sigma, h_n \models \neg \Ineq(u,t) \vee t=u.
\]
Since $n \ge \max \{\sigma(t),\sigma(u)\}$, by Lemma~\ref{lem:hn2forallhBase} (3), $\sigma, h_n \models \neg \Ineq(u,t)$ is equivalent to 
\[
\sigma, h \models H \to \neg \Ineq(u,t) \text{\quad for all $h$.}
\]
Clearly, $\sigma, h_n \models t=u$ is equivalent to 
\[
\sigma, h \models t=u \text{\quad for all $h$.}
\]
Therefore, we have 
\[
\sigma, h \models (H \to \neg \Ineq(u,t)) \vee t=u \text{\quad for all $h$,}
\] 
which is equivalent to
\[
\sigma, h \models H \to \neg \Ineq(u,t) \vee t=u \text{\quad for all $h$. }
\]

Case 2. $A \equiv \exists x \le t. B$.
Suppose 
\[
\sigma, h_n \models H \to \neg \Ineq(t,t) \vee \exists x(\Ineq(x,t) \wedge B^\circ). 
\]
Since $\sigma, h_n \models H$ and $\sigma, h_n \models \Ineq(t,t)$, we have 
\[
\sigma, h_n \models \exists x(\Ineq(x,t) \wedge B^\circ), 
\]
that is, for some $k$ 
\begin{equation}\tag{a}
\sigma[x:=k], h_n \models \Ineq(x,t) \wedge B^\circ. 
\end{equation}
We fix $h$ in order to show
\[
\sigma, h \models H \to \neg \Ineq(t,t) \vee \exists x (\Ineq(x,t) \wedge B^\circ).
\]
Assume $\sigma, h \models H$.
If $\sigma, h \models \neg \Ineq(t,t)$, the claim trivially holds.
Consider the case $\sigma, h \models \Ineq(t,t)$.
By (a), we have $\sigma[x:=k], h_n \models \Ineq(x,t)$ for some $k$.
Thus, by Lemma~\ref{lem:PA2hnBase} (3), $k \le \sigma(t)$.
By the case condition, $\sigma, h \models \Ineq(t,t)$.
Then, by Lemma~\ref{lem:H} (4), we have 
\[
\sigma[x:=k], h \models \Ineq(x,t).
\]
Moreover, since $n \ge \max(\sigma[x:=k],B)$, 
by induction hypothesis, 
we have 
\[
\sigma[x:=k], h' \models B^\circ \text{\quad  for all $h'$.}
\]
Therefore, we have 
\[
\sigma[x:=k], h \models B^\circ.
\]
Thus, we have 
\[
\sigma[x:=k], h \models \Ineq(x,t) \wedge B^\circ,
\]
that is,
\[
\sigma, h \models \exists x (\Ineq(x,t) \wedge B^\circ). 
\]

Case 3. $A \equiv \forall x \le t. B$.
Suppose
\[
\sigma, h_n \models H \to \forall x(\neg \Ineq(x,t) \vee B^\circ).
\]
Since $\sigma, h_n \models H$, we have 
\[
\sigma, h_n \models \forall x (\neg \Ineq(x,t) \vee B^\circ).
\]
We fix $h$ in order to show
\[
\sigma, h \models H \to \forall x (\neg \Ineq(x,t) \vee B^\circ). 
\]
Assume $\sigma, h \models H$.
We fix $k$ in order to show 
\[
\sigma[x:=k], h \models \neg \Ineq(x,t) \vee B^\circ. 
\]
We consider the cases for $\sigma[x:=k], h \models \Ineq(x,t)$ and $\sigma[x:=k], h \models \neg \Ineq(x,t)$ separately.

Case 3.1. The case $\sigma[x:=k], h \models \Ineq(x,t)$.
Then, there is $p$ such that 
\begin{align*}
h(p)&=2,\\
h(p+1)&=\sigma[x:=k](x)+3 = k+3, \\
h(p+2)&=\sigma[x:=k](t)+3 = \sigma(t)+3.  
\end{align*}
By Lemma~\ref{lem:H} (3), we have
$k \le \sigma(t)$.
Hence, by Lemma~\ref{lem:PA2hnBase} (3), 
we have 
\[
\sigma[x:=k], h_n \models \Ineq(x,t).
\]
Then, 
$\sigma[x:=k], h_n \models B^\circ$
must be the case.
Since $k \le \sigma(t) \le n$,  we apply the induction hypothesis to $B$
and obtain
\[
\sigma[x:=k], h' \models B^\circ \text{\quad for all $h'$.}
\]
Hence, we have 
\[
\sigma[x:=k], h \models B^\circ.
\]
Then, we have the desired result 
\[
\sigma[x:=k], h \models \neg \Ineq(x,t) \vee B^\circ.
\]

Case 3.2. If $\sigma[x:=k], h \models \neg \Ineq(x,t)$, then
\[
\sigma[x:=k], h \models \neg \Ineq(x,t) \vee B^\circ
\]
trivially holds.

Hence in both cases, we have $\sigma[x:=k], h \models \neg \Ineq(x,t) \vee B^\circ$.

Case 4. $A \equiv \exists (x = t+u) B$.
Then, $A^\circ \equiv H \to \exists x (\Add(t,u,x) \wedge B^\circ)$.
We fix $h$ and assume
\[
\sigma, h \models H
\]
 in order to show
\[
\sigma, h \models \exists x (\Add(t,u,x) \wedge B^\circ). 
\]
Since $\sigma, h_n \models H$, we have
\[
\sigma, h_n \models \exists x (\Add(t,u,x) \wedge B^\circ).
\]
That is, there exists $k$ such that 
\[
\sigma[x:=k],h_n \models \Add(t,u,x) \wedge B^\circ, 
\]
which is equivalent to 
\[
\sigma[x:=k],h_n \models \Add(t,u,x) \text{ and } \sigma[x:=k],h_n \models B^\circ. 
\]
By Lemma~\ref{lem:hn2forallhBase} (1), $\sigma[x:=k],h_n \models \Add(t,u,x)$ is equivalent to 
\[
\sigma[x:=k],h' \models H \to \Add(t,u,x) \text{\quad for all $h'$.}
\]
Since we assumed $\sigma, h \models H$, we have 
\[
\sigma[x:=k],h \models \Add(t,u,x). 
\]
Moreover, since $n \ge \max(\sigma[x:=k],B)$, by induction hypothesis for $B$, we have 
\[
\sigma[x:=k],h' \models B^\circ \text{\quad for all $h'$.}
\]
Thus, we have
\[
\sigma[x:=k],h \models B^\circ. 
\]
Therefore, we have
\[
\sigma[x:=k],h \models \Add(t,u,x) \wedge B^\circ,  
\]
that is, 
\[
\sigma, h \models \exists x(\Add(t,u,x) \wedge B^\circ). 
\]

Case 5. $A \equiv \exists(x=y\times z) B$. This case can be shown similarly to Case 4 (except it uses Lemma~\ref{lem:hn2forallhBase} (2)).
\end{proof}

Now we have the main lemma, which says that
the truth of a normal formula with $\forall$ in $\PA$
for the standard model is 
the same as the truth of its translation in $\SLN$
for the standard interpretation for all heaps.

\begin{lem} \label{lem:1}
 If $A$ is a normal formula in $\PA$, $\sigma \models \forall x A$ if and only if $\sigma, h \models (\forall x A)^\circ$ for all $h$.
\end{lem}

\begin{proof}
 $\sigma \models \forall x A$ is equivalent to 
\[
\sigma[x:=k] \models A \text{\quad for all $k \in \bN$}.
\]
We fix $k$.
Let $n \ge \max(\sigma[x:=k], A)$.
By Lemma~\ref{lem:PA2hn}, $\sigma[x:=k] \models A$ is equivalent to 
\[
\sigma[x:=k], h_n \models A^\circ.
\]
Then, by Lemma~\ref{lem:hn2forallh}, this is equivalent to $\sigma[x:=k], h \models A^\circ$ for all $h$.
Therefore, $\sigma[x:=k] \models A$ is equivalent to
\[
\sigma[x:=k], h \models A^\circ \text{\quad for all $h$.}
\]
Hence, $\sigma[x:=k] \models A$ for all $k$ is equivalent to 
\[
\sigma[x:=k], h \models A^\circ \text{\quad for all $h$ for all $k$}.
\]
Thus, $\sigma \models \forall x A$ is equivalent to
\[
\sigma, h \models \forall x A^\circ \text{\quad for all $h$,}
\]
that is, 
\begin{equation*}
\sigma, h \models (\forall x A)^\circ \text{\quad for all } h. \tag*{\qedhere}
\end{equation*}
\end{proof}

\section{Translation from $\PA$ into $\SLN$}
\label{sec:SLle2SL}

In this section, we will present
the translation of a $\Pi^0_1$ formula in $\PA$
to a formula in $\SLN$ and
prove that the translation preserves 
the validity and the non-validity.
In order to define the translation,
first we will define
a translation of a $\Pi^0_1$ formula in $\PA$ into
an equivalent normal formula with one universal quantifier in $\PA$.
Finally we will define the translation by combining the two translations
and will present the main theorem, which says
a $\Pi^0_1$ formula in $\PA$ can be simulated
in the weak fragment $\SLN$ of separation logic.
We also discuss a counterexample for the translation
when we extend it to $\Sigma^0_1$ formulas.

First we will transform
a $\Pi^0_1$ formula in $\PA$ into
a normal formula with one universal quantifier in $\PA$.
For simplicity,
we use vector notation $\overrightarrow{e}$ for a sequence $e_1, ..., e_n$ of objects.

The next proposition says that
for a given bounded formula in $\PA$ we get some equivalent normal formula.
To prove the next proposition,
we will translate a given formula
by replacing some $u+v$ or $u\times v$ by a fresh variable $z$ and
adding $\exists(z=u+v)$ or $\exists(z=u\times v)$
so that
$+$ and $\times$ appear only in the form of $\exists(z=u+v)$ or
$\exists(z=u\times v)$.
To get this, we take an innermost occurrence of $u+v$ or $u \times v$.
Moreover,
to avoid overlapping, we take a leftmost occurrence of them.

\begin{prop} \label{prop:1}
 If $A$ is a bounded formula in $\PA$, there is a normal formula $B$ such that $A \leftrightarrow B$ is valid.
\end{prop}

\begin{proof}
First, transform $A$ into a prenex normal form and
replace each occurrence of
\[
\neg (t \le u)
\]
 by
\[
u \le t \wedge u \neq t
\]
to obtain
\[
A' \equiv \overrightarrow{Qx \le t}. C, 
\]
where $C$ is a quantifier-free 
disjunctive normal form without formulas of the form $\neg (t \le u)$.
Choose the leftmost occurrence among the innermost occurrences of $u+v$
or $u \times v$ in $A'$ and explicitly denote it by $A'[u+v]$ or
$A'[u \times v]$.

Let $A'[z]$ be the formula obtained from $A'[u+v]$ or $A'[u \times v]$
by replacing the occurrence of $u+v$ or $u \times v$ in $A'$ by a fresh variable $z$.
Define
\[
\overrightarrow{Qx' \le t'}. D
\]
 by
\[
A'[z] \equiv \overrightarrow{Qx' \le t'}. D
\]
where $\overrightarrow{Q x' \le t'}$ is the longest prefix such that $z$ is not in $t'$,
namely, it has the longest $\overrightarrow {Q x' \le t'}$
among such $\overrightarrow{Q x' \le t'}$'s.
We transform $D$ into
\[
 \exists(z=u+v) D \text{ or } \exists(z=u \times v) D.
\]

We repeat this process until we have the form 
\[
\overrightarrow{\{Q x\le y, \exists(x=t)\}} A'', 
\]
where $t$ is of the form $a+b$ or $a \times b$ for some terms $a,b$ that do not contain $+$ or $\times$, and $A''$ does not contain $+,\times$ and formulas of the form $\neg(t \le u)$.
Define $B$ as this result.
\end{proof}

The prenexing and disjunctive normal form steps used to
obtain $B$ from a bounded $\PA$ formula $A$ can incur an exponential
blow-up in $|A|$. This does not affect our expressivity and
undecidability results. For the $\Pi^0_1$ completeness statement proved
in Section~\ref{sec:pi01completeness}, the upper bound argument is
independent of this blow-up. It proceeds by model-checking and
arithmetical coding rather than relying on a size-efficient translation.

We define the translation $A^\Box$ by using the proof of the previous proposition.

\begin{defi}[Translation $(\cdot )^\square$]
 Let $A \equiv \forall x B$ be a $\Pi^0_1$ formula in $\PA$, where $B$ contains only bounded quantifiers.
Let $B'$ be a normal form of $B$ obtained by the procedure described in the proof of Proposition~\ref{prop:1}.
We define $A^\square \equiv \forall x B'$.
\end{defi}

\begin{exa}
For a formula
\[
A \equiv \forall y \le x+(x+s(x)).(0 \le x+(y\times (x+y))),
\]
its translation $A^\Box$ is
\[
A^\Box \equiv \exists(x_1=x+s(x))\exists(x_2=x+x_1)\forall y \le x_2.
\\ \qquad
\exists(x_3=x+y)\exists(x_4=y\times x_3)\exists(x_5=x+x_4)(0 \le x_5).
\]
\end{exa}

Now, we have the main theorem which says that $\Pi^0_1$ formulas can be translated into $\SLN$ formulas preserving the validity and the non-validity.

\begin{thm} \label{thm:PA2SL}\label{th:represent}
 For a $\Pi^0_1$ formula $A$ in $\PA$, $A$ is valid in the standard model of $\PA$ if and only if $A^{\square \circ}$ is valid in the standard interpretation of $\SLN$.
\end{thm}

\begin{proof}
 By Proposition~\ref{prop:1} and Lemma~\ref{lem:1}.
\end{proof}

As a by-product of the above theorem, we have the undecidability of $\SLN$.

\begin{cor}\label{cor:undecidable}
 The validity of $\SLN$ formulas is undecidable.
\end{cor}

\begin{proof}
Given a Turing machine, its halting problem statement $P$ is $\Sigma^0_1$, since it can be expressed as
\[
\exists z. T(e,e,z),
\]
where $e$ is the index of the given Turing machine and $T$ is Kleene's
T-predicate which is primitive recursive (for rigorous definition, see
e.g. \cite{Shoenfield}). Thus, $\neg P$ is $\Pi^0_1$. By Theorem
\ref{thm:PA2SL}, $\neg P$ is valid in $\PA$ if and only if $(\neg
P)^{\square \circ}$ is valid in $\SLN$. If validity in $\SLN$ were
decidable, we could determine whether $P$ is true in the standard model,
contradicting the undecidability of the halting problem. Therefore,
validity in $\SLN$ is undecidable.
\end{proof}

We have just shown that $\Pi^0_1$ formulas can be translated in a way
that preserves both the validity and the non-validity. One might consider
extending the translation $(\cdot)^\circ$ by defining $(\exists x
A)^\circ \equiv \exists x A^\circ$. However, this extended translation
does not preserve the validity and the non-validity, as demonstrated in the
following proposition.

\begin{prop}\label{prop:Sigma-0-1}
 There is some $\Sigma^0_1$ closed formula $A$ such that $A$ is not valid in $\PA$ but $A^{\square \circ}$ is valid in $\SLN$.
\end{prop}

\begin{proof}
Consider the formula 
\[
A \equiv \exists x (x+0 \neq x). 
\]
This sentence is clearly not valid in $\PA$.
However, we can prove that 
\[
\sigma, h \models A^{\square \circ} \text{\quad for all $\sigma, h$}
\]
as follows.
By the procedure in the proof of Proposition~\ref{prop:1},
\[
A^\square \equiv \exists x \exists (z=x+0)(z\neq x).
\]
Thus,
\[
A^{\square \circ} \equiv \exists x (H \to \exists z(\Add(x,0,z) \wedge z \neq x)). 
\]
We fix $\sigma, h$ in order to prove
\[
\sigma,h \models A^{\square \circ}.
\]
Let
\[
n = \max \{k ~|~ h(p)=0, h(p+1)=k+3, h(p+2)=3 \} + 1
\]
and $m = n+1$. 
Let $\sigma' = \sigma[x:=n,z:=m]$.
We will show
\[
\sigma', h \models \Add(x,0,z) \wedge z \neq x
\]
assuming $\sigma', h \models H$.
By choice of $n$, we have
\[
\sigma', h \models \forall a \neg(a\hra 0, [\Bar{n}], [0]).
\]
Thus, $\sigma',h \models \Add(\Bar{n},0,\Bar{m})$ holds, because the premise of $\Add(\Bar{n},0,\Bar{m})$ is false.
Therefore, $\sigma', h \models \Add(x,0,z)$.
Furthermore, clearly $\sigma', h \models z \neq x$.
Hence, $\sigma, h \models A^{\square \circ}$ for all $h$.
\end{proof}

For the counterexample $\exists x (x+0 \neq x)$,
when a heap is given,
we can take some large argument $x$ for which
the heap is not sufficiently large.
Then $x+0$ is not computed by referring to the operation table
in the heap.
Hence the translation of $z=x+0$ becomes true.
Then by taking some $z$ different from $x$,
the translation of the counterexample becomes true.

\section{Another undecidability proof}

\def\Vars{{\rm Vars}}

In this section,
we present alternative proof of
the undecidability of validity in $\SLN$
given in Corollary~\ref{cor:undecidable},
This proof follows an approach similar to that used in \cite{ohearn01}. 
Although simpler than the proof of Theorem~\ref{th:represent}, it does 
not establish the representation of Peano arithmetic within the separation 
logic $\SLN$ with numbers.

A first-order language $L$ is defined as
that with a binary predicate symbol $P$ and
without any constants or function symbols.
Namely,
the set of terms is defined by:
\[
 t::=x,
\]
and the set of formulas is defined by:
\[
 A::= t=t \ |\ P(t,t) \ |\ \neg A \ |\ A \land A \ |\ \exists x.A.
\]

A finite structure is defined as $(U,R)$ where $U \subseteq \bN$ and $U$ is finite and $R \subseteq U^2$.
$\sigma$ is a variable assignment of $(U,R)$ if $\sigma :\Vars \to U$.
We define $\sigma_0$ as $\sigma_0(x)=0$ for all variables $x$.

We write $M,\sigma \models A$ to denote that
a formula $A$ is true by a variable assignment $\sigma$ of
a structure $M$.

The idea of this proof is to encode 
a finite structure $(U,R)$ for the language $L$ by 
a heap $h$ such that

- $n \in U$ iff $h$ has some entry of $0,n+2$, and

- $(n,m) \in R$ iff $h$ has some entry of $1,n+2,m+2$.

\begin{defi}
For a given finite structure $M=(U,R)$ of $L$,
we define the heap $h_M$ by
\[
\Dom(h_M)=\{0, 1, \ldots, 2k+3l-1 \}, \\
h_M(x)=0 \ \ (x = 2i, i<k), \\
h_M(x)=p_i+2 \ \ (x = 2i+1, i<k), \\
h_M(x)=1 \ \ (x = 2k+3i, i<l), \\
h_M(x)=n_i+2 \ \ (x = 2k+3i+1, i<l), \\
h_M(x)=m_i+2 \ \ (x = 2k+3i+2, i<l),
\]
where $U = \{p_i \ |\ i < k\}$ and $R= \{(n_i,m_i) \ |\ i < l \}$.
\end{defi}
The heap $h_M$ has information of a given structure $M$.

\begin{defi}
For a given heap $h$,
if $\sigma_0,h \models \exists ax(a \intmapsto 0,s^2(x))$,
we define a structure $M_h=(U_h,R_h)$ by
\[
U_h = \{ n \ |\ \sigma_0[x:=n], h \models \exists a(a \intmapsto 0, s^2(x)) \}, \\
R_h = \{ (n,m) \ |\ \sigma_0[x:=n,y:=m], h \models \exists a(a \intmapsto \Bar{1}, s^2(x), s^2(y)) \}.
\]
\end{defi}
The structure $M_h$ is a structure represented by a given heap $h$.

We define a translation $(\cdot )^\triangle$ from $L$ into $\SLN$.
\begin{defi}
For a formula $A$ in the language $L$,
we define the formula $A^\triangle$ in $\SLN$ by
\[
(x=y)^\triangle \equiv x=y \land \exists a(a \intmapsto 0,s^2(x)), \\
(P(x,y))^\triangle \equiv \exists a(a \intmapsto \Bar{1},s^2(x),s^2(y))
\land \exists b(b \intmapsto 0,s^2(x)) \land \exists c(c \intmapsto 0,s^2(y)),
\\
(\exists x.A)^\triangle \equiv \exists x(\exists a(a \intmapsto 0,s^2(x))
\land A^\triangle), \\
(\neg A)^\triangle \equiv \neg A^\triangle, \\
(A \land B)^\triangle \equiv A^\triangle \land B^\triangle.
\]
\end{defi}

The next is a well-known theorem for finite structures \cite{book:finitemodel}.

\begin{thm}[Trakhtenbrot]\label{th:finite-undecidable}
The validity of formulas in the language $L$ for every finite structure
is undecidable.
\end{thm}

The next lemma shows the equivalence for any formulas.

\begin{lem}\label{lemma:finite-equiv}
$M,\sigma \models A$ for all finite $M$ for all variable assignments $\sigma$ of $M$
iff
$\sigma,h \models \exists ax(a \intmapsto 0,s^2(x)) \imp
\Land_{x \in \FV(A)} \exists a(a \intmapsto 0,s^2(x)) \imp A^\triangle$
for all $h$ and all variable assignments $\sigma$.
\end{lem}

\begin{proof}
If-direction:
For a given finite structure $M$, we can construct the heap $h_M$ and
by induction on $A$
we can show that
\[
\sigma,h_M \models A^\triangle
\]
iff
\[
M,\sigma \models A,
\]
for every variable assignment $\sigma$ of $M$.

Only-if-direction:
For a given heap $h$ such that
\[
\sigma_0,h \models \exists ax(a \intmapsto 0,s^2(x)), 
\]
we can construct the finite structure $M_h$ and
by induction on $A$
we can show that
\[
M_h,\sigma \models A
\]
iff
\[
\sigma,h \models A^\triangle,
\]
for every variable assignment $\sigma$ of $M_h$.
To show the only-if-direction in the statement of the lemma by
using this claim,
from the assumption 
\[
\sigma_0,h \models \exists ax(a \intmapsto 0,s^2(x)),
\]
we have $p,q$ such that
\[
 h(p)=0, h(p+1)=q+2, 
\]
and
for a given $\sigma$ we apply this claim with 
the variable assignment $\sigma'$ of $M_h$ such that
$\sigma'(x) = \sigma(x) \ (x \in \FV(A))$ and $\sigma'(x)=q \ (\text{otherwise})$.
\end{proof}

\begin{proof}[Another Proof of Corollary \ref{cor:undecidable}]
Taking a closed formula $A$ in Lemma~\ref{lemma:finite-equiv},
we have the equivalence:
$A$ is true in all finite structure $M$
iff
$\exists ax(a \intmapsto 0,s^2(x)) \imp A^\triangle$ is valid
in the standard interpretation of $\SLN$.

By Theorem~\ref{th:finite-undecidable},
validity in $\SLN$ for the standard interpretation is undecidable.
\end{proof}

\section{$\Pi^0_1$-completeness of validity in $\SLN$}
\label{sec:pi01completeness}

In this section, we will show that the validity problem for $\SLN$ is
$\Pi^0_1$-complete. Our proof strategy involves two key steps: (1)
showing that the model-checking problem for $\SLN$ formulas is
decidable, and (2) encoding the validity of $\SLN$ formulas as a
$\Pi^0_1$ formula using (1).

From Theorem~\ref{thm:PA2SL}, the lower bound for the validity problem
in $\SLN$ follows immediately.

\begin{prop}\label{prop:hardness}
 The validity problem for $\SLN$ formulas is $\Pi^0_1$-hard.
\end{prop}

The model-checking problem for $\SLN$ formulas is defined as:
\begin{defi}[Model-checking problem for $\SLN$]
The \emph{model-checking problem} for $\SLN$ formula $A$ is to decide
whether $\sigma, h \models A$ holds for given variable assignment
$\sigma$ and heap $h$.
\end{defi}

We call the arithmetic with only 0 and $s$ \emph{successor arithmetic}.

Our main idea for proving the decidability of the model-checking problem
in $\SLN$ is to bound the search space for variable values that appear
in the intuitionistic points-to operator $\hra$. For instance, the
formula $x \hra t$ is automatically false if $x$ takes a value greater
than $\max \Dom(h)$. Similarly, $t \hra x$ is automatically false if $x$
exceeds $\max \{h(a) \mid a \in \Dom(h)\}$. This observation allows us to
restrict the search space for variables involved in the intuitionistic
points-to operator. Consequently, variables requiring unbounded search
appear only in equality formulas. In other words, quantified variables
occur solely within formulas of successor arithmetic, which is
known to be decidable because it is a fragment of Presburger arithmetic
(the decidable first-theory of addition) \cite{Presburger}.

We now proceed to formalize this idea.

\begin{defi}[Address-freeness and value-freeness]
Let $A$ be a formula of $\SLN$, and $x$ be a variable. We say that
$x$ is \emph{address-free} in $A$ if $A$ does not contain any atom of the form
$s^{n}(x) \hra t$ within the scope of the quantifier $Qx$. If every
bound variable in $A$ is address-free, then $A$ is called 
\emph{address-free}.

Similarly, we say that $x$ is \emph{value-free} in $A$ if $A$ does not contain
any atom of the form $t \hra s^{n}(x)$ within the scope of $Qx$. If
every bound variable in $A$ is value-free, then $A$ is called 
\emph{value-free}.
\end{defi}

For example, the formula $\forall x (x \hra s(y) \vee x = s(z))$ is not
address-free, since $x$ appears in an atom of the form $x \hra s(y)$
within the scope of its quantifier.

The following lemma states that if a variable is not address-free in a
formula, then there exists an equivalent formula in which the variable
is address-free.

\begin{lem}\label{lem:address-free-variable}
 Let $V$ be a finite set of variables,
 $\sigma$ be a variable assignment, $h$ be a heap,
 $Qx A$ be an $\SLN$ formula,
 and the variables in $V$ are address-free in $A$.
Then, there is a formula $A'$ such that
the variables in $V \cup \{x\}$ are address-free in $A'$ and 
$\sigma, h \models Qx A \Leftrightarrow \sigma, h \models A'$.
\end{lem}

\begin{proof}
We consider the case $Q = \exists$.
Let $M = \max \Dom(h)$.
Then,
\begin{align*}
 &\sigma, h \models \exists x A \\
\Leftrightarrow &
 \sigma[x:=0], h \models A \text{ or }
 \sigma[x:=1], h \models A \text{ or }
 \dots \text{ or }
 \sigma[x:=M], h \models A \text{ or } \\
 &\quad \sigma[x:=M+1], h \models A \text{ or }
 \sigma[x:=M+2], h \models A \text{ or }
 \dots.
\end{align*}
Let $B$ be a formula obtained from $A$ by replacing each occurrence of 
$s^n(x) \hra t$ by false.
Since $\sigma[x:=d], h \not \models s^n(x) \hra t$ for $d \ge M+1$,
we can replace each occurrence of $s^n(x) \hra t$ in $A$ be false
for $\sigma[x:=d]$.
Since the variables in $V$ are address-free in $A$, 
the variables in $V \cup \{x\}$ are address-free in $B$.
Then, 
\begin{align*}
& \sigma, h \models \exists x  A \\
\Leftrightarrow &
 \sigma[x:=0], h \models A \text{ or }
 \sigma[x:=1], h \models A \text{ or }
 \dots \text{ or }
 \sigma[x:=M], h \models A \text{ or } \\
 &\quad \sigma[x:=M+1], h \models B \text{ or }
 \sigma[x:=M+2], h \models B \text{ or }
 \dots \\
 \Leftrightarrow &
 \sigma[x:=0], h \models A \text{ or }
 \sigma[x:=1], h \models A \text{ or }
 \dots \text{ or }
 \sigma[x:=M], h \models A \text{ or } \\
 &\quad \sigma, h \models \forall x \ge M+1. B \\
\Leftrightarrow &
 \sigma, h \models A[x:=0] \text{ or }
 \sigma, h \models A[x:=\Bar 1] \text{ or }
 \dots \text{ or }
 \sigma, h \models A[x:=\Bar M] \text{ or } \\
 &\quad \sigma, h \models \forall x \ge M+1. B \\
\Leftrightarrow &
 \sigma, h \models A[x:=0] \vee A[x:=\Bar 1] \vee \dots \vee A[x:=\Bar M] \vee \forall x \ge M+1. B
\end{align*}
Clearly, the variables in $V \cup \{x\}$ are address-free in 
$A[x:=0],\ldots,A[x:=\Bar M]$.
Then, we take $A'$ to be $A[x:=0] \vee A[x:=\Bar 1] \dots \vee A[x:=\Bar M] \vee \forall x \ge M+1. B$.

The case $Q = \forall$ can be proved in a similar manner.
\end{proof}

Note that, strictly speaking, $\forall x \ge M+1. B$ is not part of the
syntax of $\SLN$. However, it can be regarded as an abbreviation of the
following formula:
\[
 \forall x (x = 0 \vee x = \Bar 1 \vee \ldots \vee x = \Bar M \vee B).
\]

\pagebreak[5] % this is so that there aren't two widow lines on the last page.

\begin{exa}
Let $\max \Dom(h) = M$.
Then, for a formula $\forall x (x \hra s(y) \vee x = s(z))$, we have the
following equivalent address-free formula:
\begin{align*}
&\forall x (x \hra s(y) \vee x = s(z)) \\
\Leftrightarrow &(0 \hra s(y) \vee 0 = s(z)) \wedge (\Bar 1 \hra s(y) \vee \Bar 1 = s(z)) \wedge 
\ldots \wedge (\Bar M \hra s(y) \vee \Bar M = s(z)) \\
\wedge &(\Bar{M+1} \hra s(y) \vee \Bar{M+1} = s(z)) \wedge (\Bar{M+2} \hra s(y) \vee \Bar{M+2} = s(z)) 
\wedge \ldots \\
\Leftrightarrow &(0 \hra s(y) \vee 0 = s(z)) \wedge (\Bar 1 \hra s(y) \vee \Bar 1 = s(z)) \wedge 
\ldots \wedge (\Bar M \hra s(y) \vee \Bar M = s(z)) \\
\wedge &(\mathrm{false} \vee \Bar{M+1} = s(z)) \wedge (\mathrm{false} \vee \Bar{M+2} = s(z)) \wedge \ldots \\
\Leftrightarrow &(0 \hra s(y) \vee 0 = s(z)) \wedge (\Bar 1 \hra s(y) \vee \Bar 1 = s(z)) \wedge 
\ldots \wedge (\Bar M \hra s(y) \vee \Bar M = s(z)) \\
\wedge &(\Bar{M+1} = s(z)) \wedge (\Bar{M+2} = s(z)) \wedge \ldots \\
\Leftrightarrow &(0 \hra s(y) \vee 0 = s(z)) \wedge (\Bar 1 \hra s(y) \vee \Bar 1 = s(z)) \wedge 
\ldots \wedge (\Bar M \hra s(y) \vee \Bar M = s(z)) \\
\wedge & \forall x \ge M+1 (x = s(z)).
\end{align*}
\end{exa}

The next lemma states that every formula $A$ has an equivalent address-free
formula.

\begin{lem}\label{lem:address-free-formula}
Let $\sigma$ be a variable assignment, $h$ be a heap, and
$A$ be a quantifier-free formula of $\SLN$.
Then, there is a formula $A'$ such that
$\sigma, h \models Q_1 x_1 \ldots Q_n x_n A
\Leftrightarrow \sigma, h \models A'$, 
where $x_1,\ldots,x_n$ are address-free in $A'$.
\end{lem}

\begin{proof}
By induction on $n$.
Suppose $n > 0$.
By induction hypothesis, there is a formula $A''$ such that
\[
 \sigma[x:=d], h \models Q_2 x_2 \ldots Q_n x_n A
\Leftrightarrow \sigma[x:=d], h \models A'',
\]
where $x_2,\ldots,x_n$ are address-free in $A''$.
Therefore, we have
\[
 \sigma,h \models Q_1 x_1 \ldots Q_n x_n A
\Leftrightarrow \sigma, h \models Q_1 x_1 A''.
\]
By Lemma~\ref{lem:address-free-variable},
there is a formula $A'$ such that
\[
 \sigma,h \models Q_1 x_1 \ldots Q_n x_n A
\Leftrightarrow \sigma, h \models A',
\]
where $x_1,\ldots,x_n$ are address-free in $A'$.
\end{proof}

In a similar manner, we obtain a corresponding result for value-free
formulas. In this case, we use the bound $M = \max \{h(a) \mid a \in
\Dom(h)\}$ when proving the next lemma similar to
Lemma~\ref{lem:address-free-formula}.

\begin{lem}\label{lem:value-free-formula}
Let $\sigma$ be a variable assignment, $h$ be a heap,
and $A$ be a quantifier-free formula of $\SLN$.
Then, there is a formula $A'$ such that
$\sigma, h \models Q_1 x_1 \ldots Q_n x_n A
\Leftrightarrow \sigma, h \models A'$, 
where $x_1,\ldots,x_n$ are value-free in $A'$.
\end{lem}

Now we will show that the model-checking for $\SLN$ is decidable.

\begin{prop}\label{prop:model-checking}
For any variable assignment $\sigma$, heap $h$, and formula $A$ of
$\SLN$, it is decidable whether $\sigma, h \models A$ holds.
\end{prop}

\begin{proof}
Without loss of generality, we can assume that $A$ is a prenex normal form.
Let $A \equiv Q_1 x_1 \ldots Q_n x_n B$, where $B$ is quantifier-free.
By Lemma~\ref{lem:address-free-formula}, there is an address-free $A'$ such that
\[
 \sigma, h \models A \Leftrightarrow \sigma, h \models A'.
\]
Let $B$ be a prenex normal form equivalent to $A'$.
By applying Lemma~\ref{lem:value-free-formula} to $B$, we have a 
value-free $B'$ such that
\[
 \sigma, h \models B \Leftrightarrow \sigma, h \models B',
\]
where $B'$ is address-free and value-free. Then, we can decide whether
$\sigma,h \models t_1 \hra t_2$ for all $t_1 \hra t_2$ in $B$, since
$t_1$ and $t_2$ are closed.
We replace each $t_1 \hra t_2$ with true or false according to the validity 
of the closed formula, and obtain a formula $C$.
Since $C$ is a formula of successor arithmetic, we can decide whether
$\sigma, h \models C$.
\end{proof}

For example, the formula $\forall x (x \hra s(y) \vee x = s(z))$ has an
equivalent address-free (and value-free) formula (here $M=\max\Dom(h)$):
\begin{align*}
&(0 \hra s(y) \vee 0 = s(z)) \wedge (\Bar 1 \hra s(y) \vee \Bar 1 = s(z)) \wedge 
\ldots (\Bar M \hra s(y) \vee \Bar M = s(z)) \\
\wedge & \forall x \ge M+1 (x = s(z)).
\end{align*}
Given $\sigma$ and $h$, we can check whether $h(i) = \sigma(y)+1$, so
we can determine whether $\sigma, h \models \Bar i \hra s(y)$ for $1 \le i \le M$.
Obviously, $\sigma, h \models \Bar i = s(y)$ can also be determined.
Furthermore, $\sigma, h \models \forall x \ge M+1 (x = s(z))$ is decidable,
because this is a formula of pure successor arithmetic (clearly, it is false).
This way, we can decide whether $\sigma, h \models \forall x (x \hra s(y) \vee x = s(z))$.

We are now ready to establish our main result: the validity problem in
$\SLN$ is $\Pi^0_1$-complete.

\begin{prop}\label{prop:pi01}
The validity problem for $\SLN$ formulas belongs to the class $\Pi^0_1$.
\end{prop}

\begin{proof}
Given variable assignment $\sigma$, heap $h$, and $\SLN$ formula $A$,
the model-checking problem $\sigma, h \models A$ is decidable by
Proposition~\ref{prop:model-checking}.
Note that the value of $\sigma$ on variables that does not appear in $A$
is irrelevant. Thus, $\sigma$ can be regarded as a finite map.
Moreover, $h$ is a finite map.
Since both the variable assignment $\sigma$ (restricted to the finitely many free variables of $A$) and heap $h$ have finite graphs, we encode them as natural numbers by a standard G\"odel coding. One convenient choice is:
\begin{align*}
 &\ulcorner \sigma \urcorner = \mathsf{Seq}(\langle v_1,\sigma(v_1)\rangle,
\ldots, \langle v_i,\sigma(v_i)\rangle), \\
 &\ulcorner h \urcorner = \mathsf{Seq}(\langle a_1,h(a_1)\rangle,
\ldots, \langle a_j,h(a_j)\rangle)
\end{align*}
where $\langle\cdot,\cdot\rangle$ is Cantor’s pairing function, the pairs 
$(v_1,\sigma(v_1)),\ldots,(v_i,\sigma(v_i))$ and $(a_1,h(a_1)),\ldots,
(a_j,h(a_j))$ enumerate the finite graphs of $\sigma$ and $h$, respectively, 
and $\mathsf{Seq}(b_1,\ldots,b_k)$ denotes the standard sequence encoding 
(for a precise definition, see, for example, \cite[Chapter 6]{Shoenfield}).
Therefore, we can express the relation $\sigma,h \models A$ as a decidable 
arithmetical predicate.
Let the predicate be $R(\sigma,h,A)$.
Then, the validity can be expressed by
\[
 \forall \sigma \forall h R(\sigma, h, A),
\]
which is a $\Pi^0_1$ formula.
\end{proof}

\begin{thm}\label{thm:pi01completeness}
 The validity problem for $\SLN$ formulas is $\Pi^0_1$-complete.
\end{thm}

\begin{proof}
By Proposition~\ref{prop:hardness} and Proposition~\ref{prop:pi01}.
\end{proof}

\section{Conclusion}
\label{sec:conclusion}

In this paper, we have shown that a minimal fragment of
separation logic—comprising only the intuitionistic points-to predicate
$\hra$, the constant 0, and the successor function—is sufficiently
expressive to simulate all $\Pi^0_1$ formulas of Peano
Arithmetic. Through a carefully constructed translation, we proved that
validity in Peano Arithmetic corresponds precisely to validity in this
fragment under the standard interpretation. This result establishes the
undecidability of validity in the fragment, despite its syntactic
simplicity.

We further showed that the validity problem in this fragment is
$\Pi^0_1$-complete by proving the decidability of model-checking and
expressing validity as a $\Pi^0_1$ formula. Additionally, we provided an
alternative undecidability proof via a reduction from finite model
theory, reinforcing the robustness of our main result.

Our findings reveal that even a highly restricted form of separation
logic can encode significant arithmetic reasoning, including properties
such as consistency and non-termination. This contributes to a deeper
understanding of the expressive boundaries of separation logic and its
interaction with arithmetic.

Future work includes exploring translations from other logical systems
into this minimal fragment, investigating whether further restrictions
or alternative arithmetic theories could yield decidable fragments, and
examining the implications of our results for automated reasoning and
program verification.

\section*{Acknowledgment}
\noindent This work was supported by JSPS KAKENHI Grant Number JP25K14999 and MEXT KAKENHI Grant Number JP25H00446. 

\bibliographystyle{alphaurl}
\bibliography{main}

\end{document}